\documentclass[aps,pra,twocolumn]{revtex4-1}

\usepackage{amsmath}
\usepackage{graphicx}
\usepackage{amsfonts}
\usepackage{cleveref}
\usepackage{ulem}

\newcommand{\eql}[2]{\begin{equation}\label{#1} #2 \end{equation}}

\newcommand{\ali}[1]{\begin{align} #1 \end{align}}
\newcommand{\<}{\langle}
\renewcommand{\>}{\rangle}

\begin{document}

\title{On collective Rabi splitting in nanolasers and nano-LEDs}

\author{Emil C. Andr\'{e}}
\email[*]{emcoa@fotonik.dtu.dk}
\affiliation{Department of Photonics Engineering, Technical University of Denmark, DK-2800 Kgs. Lyngby, Denmark}
\author{Igor E. Protsenko}
\author{Alexander V. Uskov}
\affiliation{Lebedev Physical Institute of RAS, Leninsky Prospekt 53, Moscow, 119991, Russia}
\author{Jesper M{\o}rk}
\author{Martijn Wubs}
\affiliation{Department of Photonics Engineering, Technical University of Denmark, DK-2800 Kgs. Lyngby, Denmark}

%% To be edited by editor
% \dates{Compiled \today}

%% To be edited by editor
% \doi{\url{http://dx.doi.org/10.1364/XX.XX.XXXXXX}}

\begin{abstract}
We analytically calculate the optical emission spectrum of nanolasers and nano-LEDs based on a model of many incoherently pumped two-level emitters in a cavity. At low pump rates we find  two peaks in the spectrum for large coupling strengths and numbers of emitters. We interpret the double-peaked spectrum as a signature of collective Rabi splitting, and discuss the difference between the splitting of the spectrum and the existence of two eigenmodes. We show that an LED will never exhibit a split spectrum, even though it can have distinct eigenmodes. For systems where the splitting is possible we show that the two peaks merge into a single one when the pump rate is increased. Finally, we compute the linewidth of the systems, and discuss the influence of inter-emitter correlations on the lineshape.
\end{abstract}

\maketitle

%\section{Introduction}
It is a well-established fact that the interaction of light and matter depends on the environment of the electric field and the emitters. By placing emitters in photonic cavities it is possible to increase or decrease the spontaneous emission rate \cite{Purcell}, and manipulate the emitter-field coupling strength. When this coupling strength becomes large enough to overcome the losses of the system, an excitation can be exchanged multiple times between the emitters and the field before it decays, giving rise to Rabi oscillations \cite{Allen87,Shore93} and a splitting of the cavity transmission spectrum.  This may involve a single emitter interacting with the cavity or multiple emitters acting coherently to create a collective Rabi splitting (CRS). These phenomena have been seen in transmission experiments involving several two-level atoms in a cavity \cite{Raizen} which, in the low excitation limit, are analogous to a classical resonant dispersive medium in a cavity \cite{Zhu1990}, and a single trapped atoms in a cavity \cite{Thompson,Boca04}. Rabi splitting has also been observed in micro- and nanocavities with incoherently pumped single emitters \cite{Reithmaier04,Yoshie04,Munch09,Nomura10,Gies17}, and this has been successfully modelled in e.g. \cite{Gies17, Laussy08}. Here we present a model for a large collection of incoherently pumped two-level emitters in a cavity and show that CRS can occur in this system. Our model is relevant for nanolasers and -LEDs, which have gained much interest in recent years \cite{Moelbjerg13, PAN20181, Mayer2016}  as ideal candidates for integrated light emitters for optical interconnects \cite{PAN20181} and on-chip optical circuits \cite{Mayer2016}.  We find a condition for strong coupling and collective eigenmode splitting equivalent to that in \cite{Reithmaier04,Yoshie04,Munch09,Nomura10}, 
but identify a separate condition for the optical emission spectrum to exhibit two peaks. We further find that the magnitude of the eigenmode splitting and the position of the peaks of the spectrum depend on the pump rate, and  predict a transition from strong to weak collective coupling as the pump rate increases similarly to the case of a single emitter laser~\cite{Gies17}. Importantly, we find that there is a correlation between a double-peaked spectrum and the possibility of lasing.

%In section 2 we outline basic equations and procedures. In section 3 we examine the properties of the optical emission spectrum, and discuss the conditions for observing CRS. We show that the splitting decreases with increasing pump rate, and consider the difference between eigenmode splitting and a double-peaked spectrum. In section 4 we compute the linewidth of the emission spectrum when there is only a single peak. Results are summarized and discussed in section 5. 

%\section{The model and the spectral approach}
As in \cite{Protsenko}, we consider a collection of $N_0 \gg 1$ identical two-level emitters in a single mode cavity of resonant angular frequency $\omega_0$ with no detuning between the emitters and the cavity mode. We describe the dynamics of the operators $\hat{a}$ for the electric field, $\hat{v}=\sum_{i=1}^{N_0}{\hat{v}_i}$ for the total polarization and $\hat{N}_e$, $\hat{N}_g$ for the excited and ground state populations by the Maxwell-Bloch Eqs. \cite{Scully97}
	\ali{ \dot{\hat{a}} &= \Omega_0 \hat{v} - \kappa \hat{a} + \hat{F}_{a} \label{Field}\\
	\dot{\hat{v}} &= -\frac{\gamma_\perp}{2}\hat{v} +\Omega_0 f (\hat{N}_e-\hat{N}_g)\hat{a} + \hat{F}_{v} \label{Polarization} \\
	\dot{\hat{N}}_e &= \gamma_\parallel (P \hat{N}_g - \hat{N}_e) - \Omega_0 (\hat{a}^\dagger \hat{v} + \hat{v}^\dagger \hat{a}) + \hat{F}_{N_e} \label{UpperLevel} \\
	\dot{\hat{N}}_g &= -\gamma_\parallel (P \hat{N}_g - \hat{N}_e) + \Omega_0 (\hat{a}^\dagger \hat{v} + \hat{v}^\dagger \hat{a}) + \hat{F}_{N_g}\label{LowerLevel}
    }
Here $\Omega_0$ is the vacuum Rabi frequency and $2\kappa$ is the decay rate of the cavity mode. The decay rates of the polarization and the upper level population are  $\gamma_{\perp}$ and $\gamma_{\parallel}$, respectively, and $\gamma_{\parallel}P$ is the pump rate. The factor $f$ denotes the average of squared couplings between emitters and the cavity mode, which may be combined with $\Omega_0$ to give the effective emitter-field coupling strength $g:=(\Omega_0^2 f)^{1/2}$, and $\hat{F}_{\alpha}$ is the Langevin force associated with the operator $\hat{\alpha} = \{\hat{a}, \hat{v}, \hat{N}_e,\hat{N}_g\}$. 

Using (\ref{Field})-(\ref{LowerLevel}) we found in~\cite{Protsenko} equations for the mean dynamics of the binary products of operators for the photon number $\<n\>:=
\<\hat{a}^\dagger \hat{a}\>$, population inversion $\<N\> := \<\hat{N}_e\> - \<\hat{N}_g\>$, emitter-field correlation $\<\Sigma\>:=\<\hat{a}^\dagger \hat{v}\> + \<\hat{v}^\dagger \hat{a}\>$ and inter-emitter correlations $\<D\>:=\sum_{i\neq j}
{\< \hat{v}_i^{\dagger} \hat{v}_j\>}/f$. While $\<\Sigma\>$ is often adiabatically eliminated under the assumption $\gamma_\perp \gg 2\kappa$, leading to a negligible contribution from $D$, we forego this assumption, giving four coupled equations instead of the typical two. Thus we include collective excitation trapping \cite{Leymann,Jahnke} in our model.

Here we aim to describe the steady-state spectral characteristics of the system, which were inaccessible in our previous approach, by considering the Fourier components of the operators. We will focus primarily on the region of pump power below and near threshold, where $\left<n\right> \ll N_0$ and stimulated emission plays only a small role, meaning the influence of population variations on the photon number is negligible, as in \cite{Protsenko99}. Hence, the population inversion operator $\hat{N}=\hat{N}_e-\hat{N}_g$ will be replaced by its mean value $N:=\<N\>$. From Eqs. (\ref{Field}) and (\ref{Polarization}), we then obtain the following expressions for the Fourier components of the field and polarization operators
	\ali{(\kappa-i\omega)\hat{a}(\omega) &= \Omega_0 \hat{v}(\omega) + \hat{F}_a(\omega) \label{FField1}\\
    \left(\frac{\gamma_\perp}{2} - i\omega\right)\hat{v}(\omega) &= \Omega_0 f \hat{a}(\omega)N+\hat{F}_v(\omega). \label{FPolarization1}
    }
By solving (\ref{FField1})-(\ref{FPolarization1}) for $\hat{a}(\omega)$ and $\hat{v}(\omega)$ and evaluating inverse Fourier integrals, we can then compute the steady-state photon number $\<n\> = \<\hat{a}^\dagger \hat{a}\>$ and the emitter-field correlation $\<\Sigma\> = \<\hat{a}^\dagger \hat{v}\>+\<\hat{v}^\dagger \hat{a}\>$ in terms of level populations. Upon inserting these results into the equation for the steady state population inversion $N$, we can derive analytical expressions for all three as functions of the pump rate, and these are exactly the same as we found in \cite{Protsenko} using another method. In this approach, however, we gain access to the optical emission spectrum.

\begin{figure}[t]
\centering
\includegraphics[width=\linewidth]{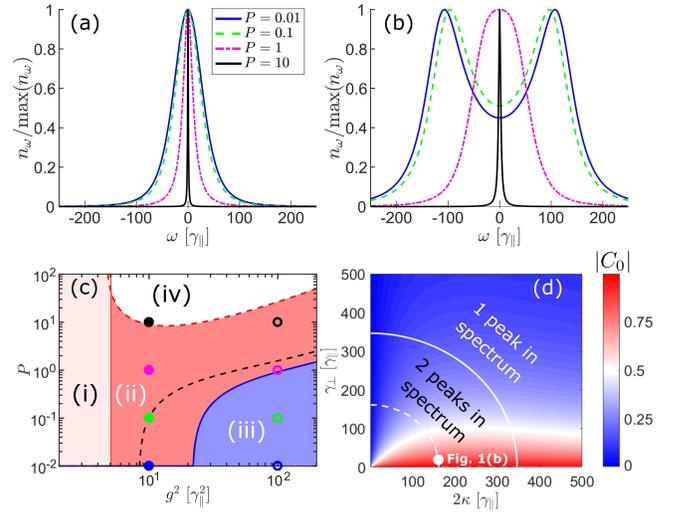}
\caption{(a),(b) Normalized emission spectrum \eqref{Spectrum1} at different values of the normalized pump rate $P$, computed using $N_0=150$ emitters, $\kappa=80 \gamma_\parallel$ and $\gamma_\perp=19 \gamma_\parallel$. (a) $g^2 = 10 \gamma_\parallel^2$, (b) $g^2 = 100 \gamma_\parallel^2$. (c) $P$ vs. squared coupling strength at the same values of $N_0$, $\kappa$ and $\gamma_\perp$. Included are graphs of $P_{\rm St}$ (red dashed), $P_{\rm c}$ (blue full) and $P_{\rm E}$ (black dashed). The points correspond to the graphs in (a) [full] and (b) [open]. (d) Parameter space $(2\kappa,\gamma_\perp)$ at $N_0$ and $g^2$ as in (b) with a contour plot of the magnitude $|C_0|$ of the inter-emitter correlations at $P\ll 1$. The curves show parameter values where the maximal splitting is constant. The point shows the parameters used in (b). 
}
\label{Fig:Spectrum}
\end{figure}

%\section{Optical Emission Spectrum}
In general, the steady-state mean intra-cavity photon number can be written as the integral over the optical emission spectrum $n_\omega$. From Eqs. (\ref{FField1})-(\ref{FPolarization1}) it can be shown that
	\eql{Spectrum1}{n_\omega = \frac{g^2 \gamma_\perp N_e}{(\omega^2-\omega_{+}^2)(\omega^2-\omega_{-}^2)},}
where the $\omega_\pm$ are given by
	\ali{\omega_\pm &:=-i\frac{(2\kappa+\gamma_\perp)}{4} \pm i\sqrt{\frac{(2\kappa-\gamma_\perp)^2}{16} + g^2 N},\label{Roots}}
Observe that the roots $\omega_\pm$ depend on the steady-state population inversion $N$, implying that if $N$ is less than 
	\eql{eigenmodes}{N_{\rm E} := -(2\kappa-\gamma_\perp)^2/(16 g^2)}
the $\omega_\pm$ have non-zero real parts. The two Lorentzians constituting the spectrum (\ref{Spectrum1}) then have central frequencies away from the cavity resonance, corresponding to two distinct coupled modes. As long as the $\omega_\pm$ are purely imaginary, the spectrum exhibits a single peak, which becomes increasingly narrow as the pump rate increases (Fig. \ref{Fig:Spectrum}(a)). However, when $\omega_\pm$ obtain sufficiently large real parts, the spectrum splits into two peaks (Fig. \ref{Fig:Spectrum}(b)). Note that the spectrum does not necessarily exhibit two peaks even if the $\omega_\pm$ have non-zero real parts: Only if $N$ is smaller than the critical inversion
	\eql{CriticalInversion}{N_{\rm c} := - \frac{4\kappa^2+\gamma_{\perp}^2}{8 g^2} = -\frac{1}{2}\left(\frac{2\kappa}{\gamma_\perp} + \frac{\gamma_\perp}{2\kappa}\right)N_{\rm th}}
will the splitting be resolved. Here $N_{\rm th} := \kappa\gamma_\perp/2g^2$ is the semi-classical threshold value of the population inversion.
When the pump rate increases, so does $N$, and therefore the two peaks in the spectrum eventually merge to one (Fig. \ref{Fig:Spectrum}(b)). 

The critical population inversion (\ref{CriticalInversion}) corresponds to a critical pump rate $P_{\rm c}$.
In Fig. \ref{Fig:Spectrum}(c), $P_{\rm c}$ is plotted as a function of $g^2$, along with the vertical line corresponding to $N_{\rm th}=N_0$ and the pump rate $P_{\rm St}$ at which the stimulated and spontaneous emission are equal. Though it is beyond the scope of this paper to discuss laser thresholds, $P_{\rm St}$ can approximately indicate the range of pump rates above which lasing is possible, so we find 
four distinct regions of interest: (i) where the coupling is so weak that $N_{\rm th} > N_0$ meaning the system will behave as an LED, i.e. it will exhibit no threshold kink in $\<n\>$ or significant decrease in linewidth;
(ii) where the coupling is strong enough that the system can achieve lasing at larger $P$, but not strong enough for the spectrum to split; (iii) where the coupling is sufficiently strong and the pump sufficiently weak, so the spectrum exhibits two peaks but not yet lasing; and (iv), where the pump rate  is strong enough for stimulated emission to dominate spontaneous emission and lasing can be achieved. Depending on our choice of parameters, region (iii) can be closer to or farther from region (i), though it can never cross the vertical limit between regions (i) and (ii). We can conclude, therefore, that a system described by this model, which cannot achieve lasing at high pump rates, will not exhibit a splitting of the spectrum at low pump rates. Conversely, a split spectrum at low pump rates implies that lasing should be possible at larger $P$.

Since the minimal population inversion is $N=-N_0$ when $P=0$, we can deduce from \eqref{CriticalInversion} that the spectrum exhibits two peaks at low pump if and only if $N_{\rm c} > -N_0$, equivalent to
	\eql{ConditionSplit}{8 g^2 N_0 > 4\kappa^2+\gamma_{\perp}^2 .}
Thus, only when the product of the number of emitters and the single emitter-field coupling strength is sufficiently large will the spectrum be split at low pump rates. This allows an interpretation of the phenomenon as a collective Rabi splitting of the spectrum, due to the strong coupling between the field and all emitters. Indeed, the system can be said to be in the collective strong-coupling regime as long as $N<N_{\rm c}$, i.e. $P < P_{\rm c}$.

In the literature, the collective strong-coupling regime is often defined by the splitting of the eigenmodes, which occurs when $N < N_{E}$. This is possible if and only if 
	\eql{ConditionSplitLorentzian2}{g \sqrt{N_0} > |2\kappa-\gamma_\perp|/4,}
which happens only for $P$ less than a certain pump rate denoted $P_{\rm E}$. In Fig.~\ref{Fig:Spectrum}(c) $P_{\rm E}$ is plotted with a dashed black line, and we see that this always lies above the full blue curve associated with $P_{\rm c}$. This means that the existence of two eigenmodes is a necessary but not a sufficient condition for CRS in the optical emission spectrum. It is worth noting that the area under the dashed black line may overlap with region~(i) in Fig.~\ref{Fig:Spectrum}(c), so it should be possible for an LED to exhibit eigenmode splitting, though it will never exhibit a double-peaked emission spectrum.

Regardless of whether one chooses to define the collective strong-coupling regime via (\ref{ConditionSplit}) or (\ref{ConditionSplitLorentzian2}), we see from Fig.~\ref{Fig:Spectrum}(c) that it is possible for the system to transition from strong to weak coupling by increasing the pump rate.

The condition (\ref{ConditionSplitLorentzian2}) is equivalent to that typically used to demarcate the region of (collective) strong coupling in theory and experiments regarding transmission spectra of  cavities containing  resonant two-level emitters \cite{Raizen,Thompson,Lugiato}, and incoherently pumped single emitters \cite{Reithmaier04,Yoshie04,Munch09,Nomura10}. In %systems as 
\cite{Munch09,Gripp,Gies17} the CRS is also seen to decrease and vanish as the intensity of the incident light is increased. It is interesting that we find a similar behaviour in %the systems described by 
our model, and that the exact same condition appears for collective eigenmode splitting, especially since our model includes emitter-emitter correlations, which lead to excitation trapping \cite{Protsenko,Jahnke,Leymann} in the intra-cavity photon number.

To examine the relative effect of the inter-emitter correlations, let us define a coherence parameter $C:= \<D\>/\<N_e\>$. In steady state this is negative at low pump rates, which is related to the excitation trapping found in \cite{Protsenko}, and it takes its minimal value $C_0$ when $P\rightarrow 0$. In Fig. \ref{Fig:Spectrum}(d), $|C_0|$ is plotted over the parameter space $(2\kappa,\gamma_\perp)$ for a fixed coupling strength $g$ and number $N_0$ of emitters equal to those used in Fig. \ref{Fig:Spectrum}(b), along with two curves which are associated with constant values of the maximal splitting $\Omega_{\rm max}$ of the spectrum: The full curve shows the $(2\kappa,\gamma_\perp)$ where $\Omega_{\rm max}$ is zero, while the dashed curve corresponds to a constant $\Omega_{\rm max}$ equal to that observed in Fig. \ref{Fig:Spectrum}(b). We see that we may obtain the same maximal splitting regardless of the ratio $2\kappa/\gamma_\perp$ of decay rates, but $|C_0|$ shows that in the bad cavity case $2\kappa/\gamma_\perp>1$ we require a larger amount inter-emitter coherence to obtain this splitting than in the good cavity case (note that the spectrum amplitude is not unchanged along these curves, so the photon number $\<n\>$ will be not be invariant). Associated with the critical pump rate $P_{\rm c}$, we find that CRS can occur only when $C< -x(1+x^2)/(1+x)^3$, where $x=2\kappa/\gamma_{\perp}$.

%\section{Linewidth}
Finally, we will discuss the case where the spectrum exhibits a single peak, i.e. when $N>N_{\rm c}$ and it makes sense to define the linewidth of the emitted light in the usual way as the FWHM of the spectrum. From \eqref{Spectrum1} we find for the linewidth
	\eql{Linewidth2}{\Delta\omega = \frac{2\kappa+\gamma_\perp}{\sqrt{2}}\left[r-1+\sqrt{r^2+(r-1)^2}\right]^{1/2},}
where 
	\eql{r}{r:= \frac{8g^2 (N_{\rm th} - N)}{(2\kappa+\gamma_\perp)^2} = \frac{N_{\rm th}-N}{N_{\rm th}-N_{\rm c}}.}
    
\begin{figure}[t]
\centering
\includegraphics[width=\linewidth]{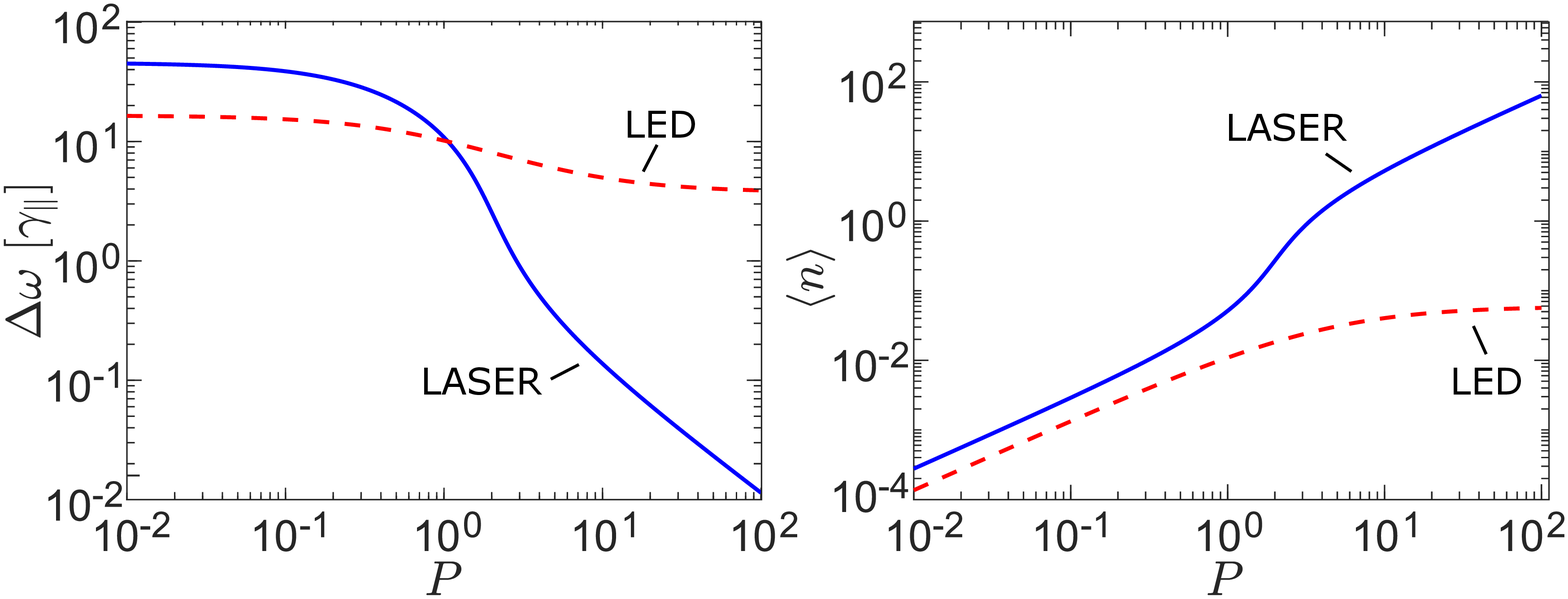}
\caption{ Linewidth (a) and photon number (b) vs. normalized pump rate for a fixed coupling strength of $g^2 =4 \gamma_{\parallel}^2$ and decay rates $(2\kappa,\gamma_\perp)=(200\gamma_{\parallel} ,10\gamma_{\parallel})$. The numbers of emitters used are $N_0=500$ for the full blue line and $N_0=100$ for the dashed red line.
}
\label{Fig:Linewidth_LED}
\end{figure}

In Fig.~\ref{Fig:Linewidth_LED}, the linewidth $\Delta\omega$ found from \eqref{Linewidth2} is shown as a function of the pump rate for two different values of $N_0$, along with the mean intra-cavity photon number $\<n\>$. For the full blue graphs, $N_0 > N_{\rm th}$, and we see a rapid decrease in $\Delta\omega$ along with a sharp increase of  $\<n\>$ as $P$ increases, in accordance with the transition to lasing. On the other hand, for the red dashed graphs the linewidth shows only a slight decrease, and the photon number saturates as the pump rate increases. This is because $N_0<N_{\rm th}$ for these graphs, whereby lasing is impossible: Indeed, the linewidth and photon number results replicate the behavior of an LED rather than a laser. 

In a system  where $N_{\rm th}<N_0$, we see that $r\ll1$ for large pump rates, and Eq.~(\ref{Linewidth2}) can then be approximated by
	\eql{Linewidth_inf}{\Delta\omega \approx \frac{2\kappa+\gamma_\perp}{2} r = \left(\frac{2\kappa\gamma_\perp}{2\kappa+\gamma_\perp}\right)^2 \frac{\hbar \omega_0}{\<P_{\rm out}\>} \frac{N_e}{N_{\rm th}}.} 
In this expression, $\<P_{\rm out}\> = \hbar\omega_0 2\kappa \<n\>$ is the mean power of the light exiting the cavity. \eqref{Linewidth_inf} is similar to expressions for the linewidth of a laser above threshold at zero detuning found elsewhere \cite{MLaxPQE66,Kuppens}, except for a factor of  1/2 which does not appear in \eqref{Linewidth_inf}. This is a known discrepancy, attributable to the neglect of population fluctuations above threshold \cite{NOEKS14_IP}: This approximation is only valid below threshold \cite{LaxCN-V,LaxCN-VI}, where the non-linearity associated with stimulated emission gives a relatively small contribution. Apart from this factor, our model thus reproduces results for the lineshape found in other models. 

%\section{Summary and Discussion}
We have considered a quantum Maxwell-Bloch equation model for nanolasers and -LEDs, and analyzed their steady-state in terms of the Fourier components of the operators for the field and polarization, including collective effects, but neglecting fluctuations of level populations. We found that the steady-state characteristics are exactly equivalent to a previous approach~\cite{Protsenko}, and we obtained an analytic expression for the emission spectrum. It was seen that the emission spectrum can behave in qualitatively different ways: When the coupling strength and number of emitters are small, the spectrum has a single peak for any pump rate, while if the coupling strength and the number of emitters becomes large enough to satisfy the inequality (\ref{ConditionSplit}), the spectrum exhibits two peaks at low pump rates. We found that the splitting decreases with increasing pump rate, and we have derived the exact values of the population inversion and the associated pump rate at which the two peaks merge into one. A similar effect was seen in transmission spectra using atoms \cite{Raizen,Zhu1990,Thompson}, where a linear, classical model is enabled by the ability to directly change the number of emitters in the cavity \cite{Zhu1990}, unlike here. We have identified the splitting of the spectrum as a collective Rabi splitting, and suggested the occurrence of this phenomenon as a way to define collective strong coupling. In conjunction with this we have discussed another possible definition of collective strong coupling in terms of eigenmode splitting, and derived an analytical condition~(\ref{ConditionSplitLorentzian2}) for this to occur. The latter condition is equivalent to that found in the literature regarding Rabi splitting in transmission spectra of many emitters in a passive cavity excited by a resonant external field~\cite{Lugiato, Raizen}. In~\cite{Laussy08,Laussy09,delValle09} the authors also suggest the splitting of the eigenmodes as the way to define whether a system is in the strong-coupling regime, but those authors, like us, deduce that this allows for a situation where the system is in the strong-coupling regime while the spectrum  only has a single peak. In particular, we have shown that it is possible for LEDs to have two distinct eigenmodes, while it is impossible for LEDs to exhibit a split emission spectrum. 

In the case where the spectrum exhibits only a single peak, we have derived an analytical expression for the width of the peak, and we have found an approximate result for the linewidth of a lasing system at high pump rates, which is similar to the reduced Schawlow-Townes linewidth found in the literature regarding the linewidth of micro- and nanolasers \cite{MLaxPQE66,Kuppens}. The integrated emission spectrum yields a mean intra-cavity photon number which agrees with that derived from the steady-state Maxwell-Bloch equations including excitation trapping, and our model describes both systems that behave like lasers, with threshold kinks in $\<n\>$ and a sudden reduction of $\Delta\omega$; as well as systems that behave like LEDs, where the photon number saturates at high pump rates, and the lineshape narrows only slightly. 

By considering the relative importance of the inter-emitter correlations, which are related to the excitation trapping observed in $\<n\>$, we found that a larger amount of coherence between the emitters is needed to see the same splitting in a system with a  poor cavity ($2\kappa \geq \gamma_\perp$) compared to one with a good cavity ($2\kappa \ll \gamma_\perp$). We can qualitatively interpret this as collective coherence helping to overcome the bad cavity to obtain CRS. This indicates a connection between the emitter cross-correlations and the collective Rabi splitting which has not been studied before which certainly merits further investigation in the future.

Our results contribute to the understanding of spectral and noise properties of quantum optical devices, and they can be expanded upon in straightforward ways: For instance, it would be interesting to include inhomogeneous broadening of the gain medium, and to connect our current model to the regime of few and single emitters. We believe that our predictions should be readily verifiable in experiments with current technology, and we would welcome experimental tests of our model. 

\section{Funding}
NATEC Centre funded by VILLUM FONDEN (grant 8692); Russian Foundation for Basic Research RFBR (Grant RFBR-17-58-150007); Russian Science
Foundation RSF (Grant 17-19-01532).

% Bibliography
\bibliography{ms}

%merlin.mbs apsrev4-1.bst 2010-07-25 4.21a (PWD, AO, DPC) hacked
%Control: key (0)
%Control: author (8) initials jnrlst
%Control: editor formatted (1) identically to author
%Control: production of article title (-1) disabled
%Control: page (0) single
%Control: year (1) truncated
%Control: production of eprint (0) enabled
\begin{thebibliography}{30}%
\makeatletter
\providecommand \@ifxundefined [1]{%
 \@ifx{#1\undefined}
}%
\providecommand \@ifnum [1]{%
 \ifnum #1\expandafter \@firstoftwo
 \else \expandafter \@secondoftwo
 \fi
}%
\providecommand \@ifx [1]{%
 \ifx #1\expandafter \@firstoftwo
 \else \expandafter \@secondoftwo
 \fi
}%
\providecommand \natexlab [1]{#1}%
\providecommand \enquote  [1]{``#1''}%
\providecommand \bibnamefont  [1]{#1}%
\providecommand \bibfnamefont [1]{#1}%
\providecommand \citenamefont [1]{#1}%
\providecommand \href@noop [0]{\@secondoftwo}%
\providecommand \href [0]{\begingroup \@sanitize@url \@href}%
\providecommand \@href[1]{\@@startlink{#1}\@@href}%
\providecommand \@@href[1]{\endgroup#1\@@endlink}%
\providecommand \@sanitize@url [0]{\catcode `\\12\catcode `\$12\catcode
  `\&12\catcode `\#12\catcode `\^12\catcode `\_12\catcode `\%12\relax}%
\providecommand \@@startlink[1]{}%
\providecommand \@@endlink[0]{}%
\providecommand \url  [0]{\begingroup\@sanitize@url \@url }%
\providecommand \@url [1]{\endgroup\@href {#1}{\urlprefix }}%
\providecommand \urlprefix  [0]{URL }%
\providecommand \Eprint [0]{\href }%
\providecommand \doibase [0]{http://dx.doi.org/}%
\providecommand \selectlanguage [0]{\@gobble}%
\providecommand \bibinfo  [0]{\@secondoftwo}%
\providecommand \bibfield  [0]{\@secondoftwo}%
\providecommand \translation [1]{[#1]}%
\providecommand \BibitemOpen [0]{}%
\providecommand \bibitemStop [0]{}%
\providecommand \bibitemNoStop [0]{.\EOS\space}%
\providecommand \EOS [0]{\spacefactor3000\relax}%
\providecommand \BibitemShut  [1]{\csname bibitem#1\endcsname}%
\let\auto@bib@innerbib\@empty
%</preamble>
\bibitem [{\citenamefont {Purcell}(1946)}]{Purcell}%
  \BibitemOpen
  \bibfield  {author} {\bibinfo {author} {\bibfnamefont {E.~M.}\ \bibnamefont
  {Purcell}},\ }\href@noop {} {\bibfield  {journal} {\bibinfo  {journal} {Phys.
  Rev.}\ }\textbf {\bibinfo {volume} {69}},\ \bibinfo {pages} {681} (\bibinfo
  {year} {1946})}\BibitemShut {NoStop}%
\bibitem [{\citenamefont {Allen}\ and\ \citenamefont {Eberly}(1987)}]{Allen87}%
  \BibitemOpen
  \bibfield  {author} {\bibinfo {author} {\bibfnamefont {L.}~\bibnamefont
  {Allen}}\ and\ \bibinfo {author} {\bibfnamefont {J.~H.}\ \bibnamefont
  {Eberly}},\ }\href@noop {} {\emph {\bibinfo {title} {Optical Resonance and
  Two-Level Atoms}}}\ (\bibinfo  {publisher} {Dover},\ \bibinfo {year}
  {1987})\BibitemShut {NoStop}%
\bibitem [{\citenamefont {Shore}\ and\ \citenamefont {Knight}(1993)}]{Shore93}%
  \BibitemOpen
  \bibfield  {author} {\bibinfo {author} {\bibfnamefont {B.~W.}\ \bibnamefont
  {Shore}}\ and\ \bibinfo {author} {\bibfnamefont {P.~L.}\ \bibnamefont
  {Knight}},\ }\href {\doibase 10.1080/09500349314551321} {\bibfield  {journal}
  {\bibinfo  {journal} {Journal of Modern Optics}\ }\textbf {\bibinfo {volume}
  {40}},\ \bibinfo {pages} {1195} (\bibinfo {year} {1993})},\ \Eprint
  {http://arxiv.org/abs/https://doi.org/10.1080/09500349314551321}
  {https://doi.org/10.1080/09500349314551321} \BibitemShut {NoStop}%
\bibitem [{\citenamefont {Raizen}\ \emph {et~al.}(1989)\citenamefont {Raizen},
  \citenamefont {Thompson}, \citenamefont {Brecha}, \citenamefont {Kimble},\
  and\ \citenamefont {Carmichael}}]{Raizen}%
  \BibitemOpen
  \bibfield  {author} {\bibinfo {author} {\bibfnamefont {M.~G.}\ \bibnamefont
  {Raizen}}, \bibinfo {author} {\bibfnamefont {R.~J.}\ \bibnamefont
  {Thompson}}, \bibinfo {author} {\bibfnamefont {R.~J.}\ \bibnamefont
  {Brecha}}, \bibinfo {author} {\bibfnamefont {H.~J.}\ \bibnamefont {Kimble}},
  \ and\ \bibinfo {author} {\bibfnamefont {H.~J.}\ \bibnamefont {Carmichael}},\
  }\href {\doibase 10.1103/PhysRevLett.63.240} {\bibfield  {journal} {\bibinfo
  {journal} {Phys. Rev. Lett.}\ }\textbf {\bibinfo {volume} {63}},\ \bibinfo
  {pages} {240} (\bibinfo {year} {1989})}\BibitemShut {NoStop}%
\bibitem [{\citenamefont {Zhu}\ \emph {et~al.}(1990)\citenamefont {Zhu},
  \citenamefont {Gauthier}, \citenamefont {Morin}, \citenamefont {Wu},
  \citenamefont {Carmichael},\ and\ \citenamefont {Mossberg}}]{Zhu1990}%
  \BibitemOpen
  \bibfield  {author} {\bibinfo {author} {\bibfnamefont {Y.}~\bibnamefont
  {Zhu}}, \bibinfo {author} {\bibfnamefont {D.~J.}\ \bibnamefont {Gauthier}},
  \bibinfo {author} {\bibfnamefont {S.~E.}\ \bibnamefont {Morin}}, \bibinfo
  {author} {\bibfnamefont {Q.}~\bibnamefont {Wu}}, \bibinfo {author}
  {\bibfnamefont {H.~J.}\ \bibnamefont {Carmichael}}, \ and\ \bibinfo {author}
  {\bibfnamefont {T.~W.}\ \bibnamefont {Mossberg}},\ }\href {\doibase
  10.1103/PhysRevLett.64.2499} {\bibfield  {journal} {\bibinfo  {journal}
  {Phys. Rev. Lett.}\ }\textbf {\bibinfo {volume} {64}},\ \bibinfo {pages}
  {2499} (\bibinfo {year} {1990})}\BibitemShut {NoStop}%
\bibitem [{\citenamefont {Thompson}\ \emph {et~al.}(1992)\citenamefont
  {Thompson}, \citenamefont {Rempe},\ and\ \citenamefont {Kimble}}]{Thompson}%
  \BibitemOpen
  \bibfield  {author} {\bibinfo {author} {\bibfnamefont {R.~J.}\ \bibnamefont
  {Thompson}}, \bibinfo {author} {\bibfnamefont {G.}~\bibnamefont {Rempe}}, \
  and\ \bibinfo {author} {\bibfnamefont {H.~J.}\ \bibnamefont {Kimble}},\
  }\href {\doibase 10.1103/PhysRevLett.68.1132} {\bibfield  {journal} {\bibinfo
   {journal} {Phys. Rev. Lett.}\ }\textbf {\bibinfo {volume} {68}},\ \bibinfo
  {pages} {1132} (\bibinfo {year} {1992})}\BibitemShut {NoStop}%
\bibitem [{\citenamefont {Boca}\ \emph {et~al.}(2004)\citenamefont {Boca},
  \citenamefont {Miller}, \citenamefont {Birnbaum}, \citenamefont {Boozer},
  \citenamefont {McKeever},\ and\ \citenamefont {Kimble}}]{Boca04}%
  \BibitemOpen
  \bibfield  {author} {\bibinfo {author} {\bibfnamefont {A.}~\bibnamefont
  {Boca}}, \bibinfo {author} {\bibfnamefont {R.}~\bibnamefont {Miller}},
  \bibinfo {author} {\bibfnamefont {K.~M.}\ \bibnamefont {Birnbaum}}, \bibinfo
  {author} {\bibfnamefont {A.~D.}\ \bibnamefont {Boozer}}, \bibinfo {author}
  {\bibfnamefont {J.}~\bibnamefont {McKeever}}, \ and\ \bibinfo {author}
  {\bibfnamefont {H.~J.}\ \bibnamefont {Kimble}},\ }\href {\doibase
  10.1103/PhysRevLett.93.233603} {\bibfield  {journal} {\bibinfo  {journal}
  {Phys. Rev. Lett.}\ }\textbf {\bibinfo {volume} {93}},\ \bibinfo {pages}
  {233603} (\bibinfo {year} {2004})}\BibitemShut {NoStop}%
\bibitem [{\citenamefont {Reithmaier}\ \emph {et~al.}(2004)\citenamefont
  {Reithmaier}, \citenamefont {Sek}, \citenamefont {L{\"o}ffler}, \citenamefont
  {Hofmann}, \citenamefont {Kuhn}, \citenamefont {Reitzenstein}, \citenamefont
  {Keldysh}, \citenamefont {Kulakovskii}, \citenamefont {Reinecke},\ and\
  \citenamefont {Forchel}}]{Reithmaier04}%
  \BibitemOpen
  \bibfield  {author} {\bibinfo {author} {\bibfnamefont {J.~P.}\ \bibnamefont
  {Reithmaier}}, \bibinfo {author} {\bibfnamefont {G.}~\bibnamefont {Sek}},
  \bibinfo {author} {\bibfnamefont {A.}~\bibnamefont {L{\"o}ffler}}, \bibinfo
  {author} {\bibfnamefont {C.}~\bibnamefont {Hofmann}}, \bibinfo {author}
  {\bibfnamefont {S.}~\bibnamefont {Kuhn}}, \bibinfo {author} {\bibfnamefont
  {S.}~\bibnamefont {Reitzenstein}}, \bibinfo {author} {\bibfnamefont {L.~V.}\
  \bibnamefont {Keldysh}}, \bibinfo {author} {\bibfnamefont {V.~D.}\
  \bibnamefont {Kulakovskii}}, \bibinfo {author} {\bibfnamefont {T.~L.}\
  \bibnamefont {Reinecke}}, \ and\ \bibinfo {author} {\bibfnamefont
  {A.}~\bibnamefont {Forchel}},\ }\href {http://dx.doi.org/10.1038/nature02969}
  {\bibfield  {journal} {\bibinfo  {journal} {Nature}\ }\textbf {\bibinfo
  {volume} {432}},\ \bibinfo {pages} {197} (\bibinfo {year}
  {2004})}\BibitemShut {NoStop}%
\bibitem [{\citenamefont {Yoshie}\ \emph {et~al.}(2004)\citenamefont {Yoshie},
  \citenamefont {Scherer}, \citenamefont {Hendrickson}, \citenamefont
  {Khitrova}, \citenamefont {Gibbs}, \citenamefont {Rupper}, \citenamefont
  {Ell}, \citenamefont {Shchekin},\ and\ \citenamefont {Deppe}}]{Yoshie04}%
  \BibitemOpen
  \bibfield  {author} {\bibinfo {author} {\bibfnamefont {T.}~\bibnamefont
  {Yoshie}}, \bibinfo {author} {\bibfnamefont {A.}~\bibnamefont {Scherer}},
  \bibinfo {author} {\bibfnamefont {J.}~\bibnamefont {Hendrickson}}, \bibinfo
  {author} {\bibfnamefont {G.}~\bibnamefont {Khitrova}}, \bibinfo {author}
  {\bibfnamefont {H.~M.}\ \bibnamefont {Gibbs}}, \bibinfo {author}
  {\bibfnamefont {G.}~\bibnamefont {Rupper}}, \bibinfo {author} {\bibfnamefont
  {C.}~\bibnamefont {Ell}}, \bibinfo {author} {\bibfnamefont {O.~B.}\
  \bibnamefont {Shchekin}}, \ and\ \bibinfo {author} {\bibfnamefont {D.~G.}\
  \bibnamefont {Deppe}},\ }\href {http://dx.doi.org/10.1038/nature03119}
  {\bibfield  {journal} {\bibinfo  {journal} {Nature}\ }\textbf {\bibinfo
  {volume} {432}},\ \bibinfo {pages} {200} (\bibinfo {year}
  {2004})}\BibitemShut {NoStop}%
\bibitem [{\citenamefont {M\"{u}nch}\ \emph {et~al.}(2009)\citenamefont
  {M\"{u}nch}, \citenamefont {Reitzenstein}, \citenamefont {Franeck},
  \citenamefont {L\"{o}ffler}, \citenamefont {Heindel}, \citenamefont
  {H\"{o}fling}, \citenamefont {Worschech},\ and\ \citenamefont
  {Forchel}}]{Munch09}%
  \BibitemOpen
  \bibfield  {author} {\bibinfo {author} {\bibfnamefont {S.}~\bibnamefont
  {M\"{u}nch}}, \bibinfo {author} {\bibfnamefont {S.}~\bibnamefont
  {Reitzenstein}}, \bibinfo {author} {\bibfnamefont {P.}~\bibnamefont
  {Franeck}}, \bibinfo {author} {\bibfnamefont {A.}~\bibnamefont
  {L\"{o}ffler}}, \bibinfo {author} {\bibfnamefont {T.}~\bibnamefont
  {Heindel}}, \bibinfo {author} {\bibfnamefont {S.}~\bibnamefont
  {H\"{o}fling}}, \bibinfo {author} {\bibfnamefont {L.}~\bibnamefont
  {Worschech}}, \ and\ \bibinfo {author} {\bibfnamefont {A.}~\bibnamefont
  {Forchel}},\ }\href {\doibase 10.1364/OE.17.012821} {\bibfield  {journal}
  {\bibinfo  {journal} {Opt. Express}\ }\textbf {\bibinfo {volume} {17}},\
  \bibinfo {pages} {12821} (\bibinfo {year} {2009})}\BibitemShut {NoStop}%
\bibitem [{\citenamefont {Nomura}\ \emph {et~al.}(2010)\citenamefont {Nomura},
  \citenamefont {Kumagai}, \citenamefont {Iwamoto}, \citenamefont {Ota},\ and\
  \citenamefont {Arakawa}}]{Nomura10}%
  \BibitemOpen
  \bibfield  {author} {\bibinfo {author} {\bibfnamefont {M.}~\bibnamefont
  {Nomura}}, \bibinfo {author} {\bibfnamefont {N.}~\bibnamefont {Kumagai}},
  \bibinfo {author} {\bibfnamefont {S.}~\bibnamefont {Iwamoto}}, \bibinfo
  {author} {\bibfnamefont {Y.}~\bibnamefont {Ota}}, \ and\ \bibinfo {author}
  {\bibfnamefont {Y.}~\bibnamefont {Arakawa}},\ }\href
  {http://dx.doi.org/10.1038/nphys1518} {\bibfield  {journal} {\bibinfo
  {journal} {Nature Physics}\ }\textbf {\bibinfo {volume} {6}},\ \bibinfo
  {pages} {279} (\bibinfo {year} {2010})}\BibitemShut {NoStop}%
\bibitem [{\citenamefont {Gies}\ \emph {et~al.}(2017)\citenamefont {Gies},
  \citenamefont {Gericke}, \citenamefont {Gartner}, \citenamefont {Holzinger},
  \citenamefont {Hopfmann}, \citenamefont {Heindel}, \citenamefont {Wolters},
  \citenamefont {Schneider}, \citenamefont {Florian}, \citenamefont {Jahnke},
  \citenamefont {H\"ofling}, \citenamefont {Kamp},\ and\ \citenamefont
  {Reitzenstein}}]{Gies17}%
  \BibitemOpen
  \bibfield  {author} {\bibinfo {author} {\bibfnamefont {C.}~\bibnamefont
  {Gies}}, \bibinfo {author} {\bibfnamefont {F.}~\bibnamefont {Gericke}},
  \bibinfo {author} {\bibfnamefont {P.}~\bibnamefont {Gartner}}, \bibinfo
  {author} {\bibfnamefont {S.}~\bibnamefont {Holzinger}}, \bibinfo {author}
  {\bibfnamefont {C.}~\bibnamefont {Hopfmann}}, \bibinfo {author}
  {\bibfnamefont {T.}~\bibnamefont {Heindel}}, \bibinfo {author} {\bibfnamefont
  {J.}~\bibnamefont {Wolters}}, \bibinfo {author} {\bibfnamefont
  {C.}~\bibnamefont {Schneider}}, \bibinfo {author} {\bibfnamefont
  {M.}~\bibnamefont {Florian}}, \bibinfo {author} {\bibfnamefont
  {F.}~\bibnamefont {Jahnke}}, \bibinfo {author} {\bibfnamefont
  {S.}~\bibnamefont {H\"ofling}}, \bibinfo {author} {\bibfnamefont
  {M.}~\bibnamefont {Kamp}}, \ and\ \bibinfo {author} {\bibfnamefont
  {S.}~\bibnamefont {Reitzenstein}},\ }\href {\doibase
  10.1103/PhysRevA.96.023806} {\bibfield  {journal} {\bibinfo  {journal} {Phys.
  Rev. A}\ }\textbf {\bibinfo {volume} {96}},\ \bibinfo {pages} {023806}
  (\bibinfo {year} {2017})}\BibitemShut {NoStop}%
\bibitem [{\citenamefont {Laussy}\ \emph {et~al.}(2008)\citenamefont {Laussy},
  \citenamefont {del Valle},\ and\ \citenamefont {Tejedor}}]{Laussy08}%
  \BibitemOpen
  \bibfield  {author} {\bibinfo {author} {\bibfnamefont {F.~P.}\ \bibnamefont
  {Laussy}}, \bibinfo {author} {\bibfnamefont {E.}~\bibnamefont {del Valle}}, \
  and\ \bibinfo {author} {\bibfnamefont {C.}~\bibnamefont {Tejedor}},\ }\href
  {\doibase 10.1103/PhysRevLett.101.083601} {\bibfield  {journal} {\bibinfo
  {journal} {Phys. Rev. Lett.}\ }\textbf {\bibinfo {volume} {101}},\ \bibinfo
  {pages} {083601} (\bibinfo {year} {2008})}\BibitemShut {NoStop}%
\bibitem [{\citenamefont {Moelbjerg}\ \emph {et~al.}(2013)\citenamefont
  {Moelbjerg}, \citenamefont {Kaer}, \citenamefont {Lorke}, \citenamefont
  {Tromborg},\ and\ \citenamefont {M{\o}rk}}]{Moelbjerg13}%
  \BibitemOpen
  \bibfield  {author} {\bibinfo {author} {\bibfnamefont {A.}~\bibnamefont
  {Moelbjerg}}, \bibinfo {author} {\bibfnamefont {P.}~\bibnamefont {Kaer}},
  \bibinfo {author} {\bibfnamefont {M.}~\bibnamefont {Lorke}}, \bibinfo
  {author} {\bibfnamefont {B.}~\bibnamefont {Tromborg}}, \ and\ \bibinfo
  {author} {\bibfnamefont {J.}~\bibnamefont {M{\o}rk}},\ }\href {\doibase
  10.1109/JQE.2013.2282464} {\bibfield  {journal} {\bibinfo  {journal} {IEEE
  Journal of Quantum Electronics}\ }\textbf {\bibinfo {volume} {49}},\ \bibinfo
  {pages} {945} (\bibinfo {year} {2013})}\BibitemShut {NoStop}%
\bibitem [{\citenamefont {Pan}\ \emph {et~al.}(2018)\citenamefont {Pan},
  \citenamefont {Deka}, \citenamefont {Amili}, \citenamefont {Gu},\ and\
  \citenamefont {Fainman}}]{PAN20181}%
  \BibitemOpen
  \bibfield  {author} {\bibinfo {author} {\bibfnamefont {S.~H.}\ \bibnamefont
  {Pan}}, \bibinfo {author} {\bibfnamefont {S.~S.}\ \bibnamefont {Deka}},
  \bibinfo {author} {\bibfnamefont {A.~E.}\ \bibnamefont {Amili}}, \bibinfo
  {author} {\bibfnamefont {Q.}~\bibnamefont {Gu}}, \ and\ \bibinfo {author}
  {\bibfnamefont {Y.}~\bibnamefont {Fainman}},\ }\href {\doibase
  https://doi.org/10.1016/j.pquantelec.2018.05.001} {\bibfield  {journal}
  {\bibinfo  {journal} {Progress in Quantum Electronics}\ }\textbf {\bibinfo
  {volume} {59}},\ \bibinfo {pages} {1 } (\bibinfo {year} {2018})}\BibitemShut
  {NoStop}%
\bibitem [{\citenamefont {Mayer}\ \emph {et~al.}(2016)\citenamefont {Mayer},
  \citenamefont {Janker}, \citenamefont {Loitsch}, \citenamefont {Treu},
  \citenamefont {Kostenbader}, \citenamefont {Lichtmannecker}, \citenamefont
  {Reichert}, \citenamefont {Morkötter}, \citenamefont {Kaniber},
  \citenamefont {Abstreiter}, \citenamefont {Gies}, \citenamefont
  {Koblmüller},\ and\ \citenamefont {Finley}}]{Mayer2016}%
  \BibitemOpen
  \bibfield  {author} {\bibinfo {author} {\bibfnamefont {B.}~\bibnamefont
  {Mayer}}, \bibinfo {author} {\bibfnamefont {L.}~\bibnamefont {Janker}},
  \bibinfo {author} {\bibfnamefont {B.}~\bibnamefont {Loitsch}}, \bibinfo
  {author} {\bibfnamefont {J.}~\bibnamefont {Treu}}, \bibinfo {author}
  {\bibfnamefont {T.}~\bibnamefont {Kostenbader}}, \bibinfo {author}
  {\bibfnamefont {S.}~\bibnamefont {Lichtmannecker}}, \bibinfo {author}
  {\bibfnamefont {T.}~\bibnamefont {Reichert}}, \bibinfo {author}
  {\bibfnamefont {S.}~\bibnamefont {Morkötter}}, \bibinfo {author}
  {\bibfnamefont {M.}~\bibnamefont {Kaniber}}, \bibinfo {author} {\bibfnamefont
  {G.}~\bibnamefont {Abstreiter}}, \bibinfo {author} {\bibfnamefont
  {C.}~\bibnamefont {Gies}}, \bibinfo {author} {\bibfnamefont {G.}~\bibnamefont
  {Koblmüller}}, \ and\ \bibinfo {author} {\bibfnamefont {J.~J.}\ \bibnamefont
  {Finley}},\ }\href {\doibase 10.1021/acs.nanolett.5b03404} {\bibfield
  {journal} {\bibinfo  {journal} {Nano Letters}\ }\textbf {\bibinfo {volume}
  {16}},\ \bibinfo {pages} {152} (\bibinfo {year} {2016})},\ \Eprint
  {http://arxiv.org/abs/https://doi.org/10.1021/acs.nanolett.5b03404}
  {https://doi.org/10.1021/acs.nanolett.5b03404} \BibitemShut {NoStop}%
\bibitem [{\citenamefont {Protsenko}\ \emph {et~al.}()\citenamefont
  {Protsenko}, \citenamefont {Andr{\'e}}, \citenamefont {Uskov}, \citenamefont
  {M{\o}rk},\ and\ \citenamefont {Wubs}}]{Protsenko}%
  \BibitemOpen
  \bibfield  {author} {\bibinfo {author} {\bibfnamefont {I.~E.}\ \bibnamefont
  {Protsenko}}, \bibinfo {author} {\bibfnamefont {E.}~\bibnamefont
  {Andr{\'e}}}, \bibinfo {author} {\bibfnamefont {A.~V.}\ \bibnamefont
  {Uskov}}, \bibinfo {author} {\bibfnamefont {J.}~\bibnamefont {M{\o}rk}}, \
  and\ \bibinfo {author} {\bibfnamefont {M.}~\bibnamefont {Wubs}},\ }\href@noop
  {} {}\bibinfo {howpublished}
  {\url{https://arxiv.org/abs/1709.08200}}\BibitemShut {NoStop}%
\bibitem [{\citenamefont {Scully}\ and\ \citenamefont
  {Zubairy}(1997)}]{Scully97}%
  \BibitemOpen
  \bibfield  {author} {\bibinfo {author} {\bibfnamefont {M.~O.}\ \bibnamefont
  {Scully}}\ and\ \bibinfo {author} {\bibfnamefont {M.~S.}\ \bibnamefont
  {Zubairy}},\ }\href {\doibase 10.1017/CBO9780511813993} {\emph {\bibinfo
  {title} {Quantum Optics}}}\ (\bibinfo  {publisher} {Cambridge University
  Press},\ \bibinfo {year} {1997})\BibitemShut {NoStop}%
\bibitem [{\citenamefont {Leymann}\ \emph {et~al.}(2015)\citenamefont
  {Leymann}, \citenamefont {Foerster}, \citenamefont {Jahnke}, \citenamefont
  {Wiersig},\ and\ \citenamefont {Gies}}]{Leymann}%
  \BibitemOpen
  \bibfield  {author} {\bibinfo {author} {\bibfnamefont {H.~A.~M.}\
  \bibnamefont {Leymann}}, \bibinfo {author} {\bibfnamefont {A.}~\bibnamefont
  {Foerster}}, \bibinfo {author} {\bibfnamefont {F.}~\bibnamefont {Jahnke}},
  \bibinfo {author} {\bibfnamefont {J.}~\bibnamefont {Wiersig}}, \ and\
  \bibinfo {author} {\bibfnamefont {C.}~\bibnamefont {Gies}},\ }\href {\doibase
  10.1103/PhysRevApplied.4.044018} {\bibfield  {journal} {\bibinfo  {journal}
  {Phys. Rev. Applied}\ }\textbf {\bibinfo {volume} {4}},\ \bibinfo {pages}
  {044018} (\bibinfo {year} {2015})}\BibitemShut {NoStop}%
\bibitem [{\citenamefont {Jahnke}\ \emph {et~al.}(2016)\citenamefont {Jahnke},
  \citenamefont {Gies}, \citenamefont {A{\ss}mann}, \citenamefont {Bayer},
  \citenamefont {Leymann}, \citenamefont {Foerster}, \citenamefont {Wiersig},
  \citenamefont {Schneider}, \citenamefont {Kamp},\ and\ \citenamefont
  {H{\"o}fling}}]{Jahnke}%
  \BibitemOpen
  \bibfield  {author} {\bibinfo {author} {\bibfnamefont {F.}~\bibnamefont
  {Jahnke}}, \bibinfo {author} {\bibfnamefont {C.}~\bibnamefont {Gies}},
  \bibinfo {author} {\bibfnamefont {M.}~\bibnamefont {A{\ss}mann}}, \bibinfo
  {author} {\bibfnamefont {M.}~\bibnamefont {Bayer}}, \bibinfo {author}
  {\bibfnamefont {H.~A.~M.}\ \bibnamefont {Leymann}}, \bibinfo {author}
  {\bibfnamefont {A.}~\bibnamefont {Foerster}}, \bibinfo {author}
  {\bibfnamefont {J.}~\bibnamefont {Wiersig}}, \bibinfo {author} {\bibfnamefont
  {C.}~\bibnamefont {Schneider}}, \bibinfo {author} {\bibfnamefont
  {M.}~\bibnamefont {Kamp}}, \ and\ \bibinfo {author} {\bibfnamefont
  {S.}~\bibnamefont {H{\"o}fling}},\ }\href
  {http://dx.doi.org/10.1038/ncomms11540} {\bibfield  {journal} {\bibinfo
  {journal} {Nature Communications}\ }\textbf {\bibinfo {volume} {7}} (\bibinfo
  {year} {2016})}\BibitemShut {NoStop}%
\bibitem [{\citenamefont {Protsenko}\ \emph {et~al.}(1999)\citenamefont
  {Protsenko}, \citenamefont {Domokos}, \citenamefont {Lef\`evre-Seguin},
  \citenamefont {Hare}, \citenamefont {Raimond},\ and\ \citenamefont
  {Davidovich}}]{Protsenko99}%
  \BibitemOpen
  \bibfield  {author} {\bibinfo {author} {\bibfnamefont {I.}~\bibnamefont
  {Protsenko}}, \bibinfo {author} {\bibfnamefont {P.}~\bibnamefont {Domokos}},
  \bibinfo {author} {\bibfnamefont {V.}~\bibnamefont {Lef\`evre-Seguin}},
  \bibinfo {author} {\bibfnamefont {J.}~\bibnamefont {Hare}}, \bibinfo {author}
  {\bibfnamefont {J.~M.}\ \bibnamefont {Raimond}}, \ and\ \bibinfo {author}
  {\bibfnamefont {L.}~\bibnamefont {Davidovich}},\ }\href {\doibase
  10.1103/PhysRevA.59.1667} {\bibfield  {journal} {\bibinfo  {journal} {Phys.
  Rev. A}\ }\textbf {\bibinfo {volume} {59}},\ \bibinfo {pages} {1667}
  (\bibinfo {year} {1999})}\BibitemShut {NoStop}%
\bibitem [{\citenamefont {Lugiato}(1984)}]{Lugiato}%
  \BibitemOpen
  \bibfield  {author} {\bibinfo {author} {\bibfnamefont {L.~A.}\ \bibnamefont
  {Lugiato}}\ }(\bibinfo  {publisher} {Elsevier},\ \bibinfo {year} {1984})\
  pp.\ \bibinfo {pages} {69 -- 216}\BibitemShut {NoStop}%
\bibitem [{\citenamefont {Gripp}\ \emph {et~al.}(1997)\citenamefont {Gripp},
  \citenamefont {Mielke},\ and\ \citenamefont {Orozco}}]{Gripp}%
  \BibitemOpen
  \bibfield  {author} {\bibinfo {author} {\bibfnamefont {J.}~\bibnamefont
  {Gripp}}, \bibinfo {author} {\bibfnamefont {S.~L.}\ \bibnamefont {Mielke}}, \
  and\ \bibinfo {author} {\bibfnamefont {L.~A.}\ \bibnamefont {Orozco}},\
  }\href {\doibase 10.1103/PhysRevA.56.3262} {\bibfield  {journal} {\bibinfo
  {journal} {Phys. Rev. A}\ }\textbf {\bibinfo {volume} {56}},\ \bibinfo
  {pages} {3262} (\bibinfo {year} {1997})}\BibitemShut {NoStop}%
\bibitem [{\citenamefont {Lax}()}]{MLaxPQE66}%
  \BibitemOpen
  \bibfield  {author} {\bibinfo {author} {\bibfnamefont {M.}~\bibnamefont
  {Lax}},\ }\href@noop {} {\enquote {\bibinfo {title} {Physics of quantum
  electronics},}\ }\bibinfo {note} {Edited by P.L. Kelley, B. Lax and P.E.
  Tannenwald (McGraw-Hill, New York, 1966), p. 735}\BibitemShut {NoStop}%
\bibitem [{\citenamefont {Kuppens}\ \emph {et~al.}(1994)\citenamefont
  {Kuppens}, \citenamefont {van Exter},\ and\ \citenamefont
  {Woerdman}}]{Kuppens}%
  \BibitemOpen
  \bibfield  {author} {\bibinfo {author} {\bibfnamefont {S.~J.~M.}\
  \bibnamefont {Kuppens}}, \bibinfo {author} {\bibfnamefont {M.~P.}\
  \bibnamefont {van Exter}}, \ and\ \bibinfo {author} {\bibfnamefont {J.~P.}\
  \bibnamefont {Woerdman}},\ }\href {\doibase 10.1103/PhysRevLett.72.3815}
  {\bibfield  {journal} {\bibinfo  {journal} {Phys. Rev. Lett.}\ }\textbf
  {\bibinfo {volume} {72}},\ \bibinfo {pages} {3815} (\bibinfo {year}
  {1994})}\BibitemShut {NoStop}%
\bibitem [{\citenamefont {Protsenko}\ \emph {et~al.}(2018)\citenamefont
  {Protsenko}, \citenamefont {Andr{\'e}}, \citenamefont {Uskov}, \citenamefont
  {Wubs},\ and\ \citenamefont {M{\o}rk}}]{NOEKS14_IP}%
  \BibitemOpen
  \bibfield  {author} {\bibinfo {author} {\bibfnamefont {I.}~\bibnamefont
  {Protsenko}}, \bibinfo {author} {\bibfnamefont {E.}~\bibnamefont
  {Andr{\'e}}}, \bibinfo {author} {\bibfnamefont {A.}~\bibnamefont {Uskov}},
  \bibinfo {author} {\bibfnamefont {M.}~\bibnamefont {Wubs}}, \ and\ \bibinfo
  {author} {\bibfnamefont {J.}~\bibnamefont {M{\o}rk}},\ }\href@noop {}
  {\bibfield  {journal} {\bibinfo  {journal} {Proceedings of NOEKS14}\ ,\
  \bibinfo {pages} {62}} (\bibinfo {year} {2018})}\BibitemShut {NoStop}%
\bibitem [{\citenamefont {Lax}(1967)}]{LaxCN-V}%
  \BibitemOpen
  \bibfield  {author} {\bibinfo {author} {\bibfnamefont {M.}~\bibnamefont
  {Lax}},\ }\href {\doibase 10.1103/PhysRev.160.290} {\bibfield  {journal}
  {\bibinfo  {journal} {Phys. Rev.}\ }\textbf {\bibinfo {volume} {160}},\
  \bibinfo {pages} {290} (\bibinfo {year} {1967})}\BibitemShut {NoStop}%
\bibitem [{\citenamefont {Hempstead}\ and\ \citenamefont
  {Lax}(1967)}]{LaxCN-VI}%
  \BibitemOpen
  \bibfield  {author} {\bibinfo {author} {\bibfnamefont {R.~D.}\ \bibnamefont
  {Hempstead}}\ and\ \bibinfo {author} {\bibfnamefont {M.}~\bibnamefont
  {Lax}},\ }\href {\doibase 10.1103/PhysRev.161.350} {\bibfield  {journal}
  {\bibinfo  {journal} {Phys. Rev.}\ }\textbf {\bibinfo {volume} {161}},\
  \bibinfo {pages} {350} (\bibinfo {year} {1967})}\BibitemShut {NoStop}%
\bibitem [{\citenamefont {Laussy}\ \emph {et~al.}(2009)\citenamefont {Laussy},
  \citenamefont {del Valle},\ and\ \citenamefont {Tejedor}}]{Laussy09}%
  \BibitemOpen
  \bibfield  {author} {\bibinfo {author} {\bibfnamefont {F.~P.}\ \bibnamefont
  {Laussy}}, \bibinfo {author} {\bibfnamefont {E.}~\bibnamefont {del Valle}}, \
  and\ \bibinfo {author} {\bibfnamefont {C.}~\bibnamefont {Tejedor}},\ }\href
  {\doibase 10.1103/PhysRevB.79.235325} {\bibfield  {journal} {\bibinfo
  {journal} {Phys. Rev. B}\ }\textbf {\bibinfo {volume} {79}},\ \bibinfo
  {pages} {235325} (\bibinfo {year} {2009})}\BibitemShut {NoStop}%
\bibitem [{\citenamefont {del Valle}\ \emph {et~al.}(2009)\citenamefont {del
  Valle}, \citenamefont {Laussy},\ and\ \citenamefont {Tejedor}}]{delValle09}%
  \BibitemOpen
  \bibfield  {author} {\bibinfo {author} {\bibfnamefont {E.}~\bibnamefont {del
  Valle}}, \bibinfo {author} {\bibfnamefont {F.~P.}\ \bibnamefont {Laussy}}, \
  and\ \bibinfo {author} {\bibfnamefont {C.}~\bibnamefont {Tejedor}},\ }\href
  {\doibase 10.1103/PhysRevB.79.235326} {\bibfield  {journal} {\bibinfo
  {journal} {Phys. Rev. B}\ }\textbf {\bibinfo {volume} {79}},\ \bibinfo
  {pages} {235326} (\bibinfo {year} {2009})}\BibitemShut {NoStop}%
\end{thebibliography}%

\end{document}